# Ceramic processing and multiferroic properties of the perovskite YMnO$_3$ – BiFeO$_3$ binary system

Jose A. Quintana-Cilleruelo[1,*], Alicia Castro[1], Harvey Amorín[1],

Vignaswaran K. Veerapandiyan[2], Marco Deluca[2], Octavio Peña[3] and Miguel Algueró[1]

[1] Instituto de Ciencia de Materiales de Madrid, CSIC. Cantoblanco, 28049 Madrid, Spain
[2] Materials Center Leoben Forschung GmbH, Roseggerstr. 12, 8700 Leoben, Austria
[3] Sciences Chimiques de Rennes, UMR 6226, Université de Rennes 1, 35042 Rennes, France.

The perovskite (1-x)YMnO$_3$-xBiFeO$_3$ binary system is very promising because of its multiferroic end members. Nanocrystalline phases have been recently obtained by mechanosynthesis across the system, and the perovskite structural evolution has been described. Two continuous solid solutions with orthorhombic *Pnma* and rhombohedral *R*3*c* symmetries were found, which coexist within a broad compositional interval of 0.5≤x≤0.9. This might be a polar-nonpolar morphotropic phase boundary region, at which strong phase-change magnetoelectric responses can be expected. A major issue is phase decomposition at moderate temperatures that highly complicates ceramic processing. This is required for carrying out a sound electrical characterization and also for their use in devices. We present here the application of Spark Plasma Sintering to the ceramic processing of YMnO$_3$-BiFeO$_3$ phases. This advanced technique, when combined with nanocrystalline powders, allowed densifying phases at reduced processing temperatures and times, so that perovskite decomposition was avoided. Electrical measurements were accomplished, and the response was shown to be mostly dominated by conduction. Nonetheless, the intrinsic dielectric permittivity was obtained, and a distinctive enhancement in the phase coexistence region was revealed. Besides, Rayleigh-type behaviour characteristic of ferroelectrics was also demonstrated for all rhombohedral phases. Magnetic characterization was performed in this region, and antiferromagnetism was shown.

---

[*] Corresponding autor: joseangel.quintana@csic.es

Introduction

The search for single-phase multiferroics with large magnetoelectric (ME) response at room temperature is driven by the perspective of enabling a range of potentially disruptive technologies, like electrically tunable magnetic microwave components, spintronic devices and nonvolatile magnetoelectric memories, among others.[1,2] One of the most promising approaches for the design of new multiferroics is the chemical engineering of $ABO_3$ perovskite oxides. This approach is based on the stereochemical activity of a lone pair cation at the A-site (often $Bi^{3+}$) to induce an off-center structural distortion and ferroelectricity, and on a magnetically active cation at the B-site to cause magnetic ordering[3]. This is the case of $BiFeO_3$, the most topical multiferroic material due to its large spontaneous polarization (~100 µC·cm$^{-2}$),[4] antiferromagnetism and high ferroic ordering temperatures (Curie temperature $T_C$ = 825°C, Neel temperature $T_N$ = 370°C).[5] Therefore, $BiFeO_3$ presents multiferroicity at room temperature, though the extremely high coercive field and conductivity, along with its antiferromagnetic nature, hinder its application in devices.[5] Besides, perovskite single-phase materials are difficult to prepare, and secondary phases like mullite $Bi_2Fe_4O_9$ and sillenite $Bi_{25}FeO_{39}$ are often present in $BiFeO_3$.[6,7] As an alternative, $BiFeO_3$ modifications are being extensively investigated, and materials with tailored electrical properties and weak ferromagnetism have been obtained free of the secondary phases.[8,9]

Another highly topical multiferroic material is $YMnO_3$, a geometric ferroelectric with hexagonal structure that develops antiferromagnetic order at ~70 K.[10] $YMnO_3$ also presents a high-density perovskite orthorhombic phase (*Pnma*, S. G. No 62), also an improper ferroelectric but of magnetic origin and then, multiferroic after a rearrangement of the antiferromagnetic order at 28 K.[11] This orthorhombic phase can be obtained at high pressure[12] or by other non-conventional methods, like soft chemistry in oxidizing media (resulting in a final mixture of hexagonal and orthorhombic YMnO$_3$)[13] and mechanosynthesis.[14] This latter technique, which promotes the synthesis of high-density phase under the high point pressures of the order of GPa exerted during the milling,[15] is an environmentally friendly, up-scalable alternative to high-pressure synthesis that has allowed obtaining perovskite single-phase materials free of the YMnO$_3$ hexagonal phase.[14, 16] It involves mechanical milling that is employed not only to amorphize oxide materials and generate physical changes (mechanoactivation), but also to induce chemical transformations and synthetize new species. Mechanosynthesis provides nanoscale particle size and chemical homogeneity, and highly facilitates densification and phase control during sintering.[17, 18]

Perovskite binary systems involving end members with different structural symmetry can lead to the appearance of a region in the phase diagram where two or more polymorphs with the same condition coexist, known as a Morphotropic Phase Boundary (MPB).[19] Enhancement of electrical and electromechanical properties generally takes place at MPBs if both phases are ferroelectric, and the possibility of obtaining multiferroic MPBs that show enhanced magnetoelectric coefficients is being actively investigated.[20, 21] In this case, both ferroelectric polymorphic forms envolve from the same perovskite non-polar parent phase. An alternative type of MPBs are polar/non-polar ones involving ferroelectric and paraelectric polymorphs, at which property enhancement by a polarization extension mechanism is expected.[22, 23] Rhombohedral *R3c*-Orthorhombic *Pnma* examples have been found in rare-earth substituted BiFeO$_3$ (Bi$_{1-x}$RE$_x$FeO$_3$ where RE = Dy, Gd, Sm).[24] Note that orthorhombic YMnO$_3$ is not ferroelectric and has the *Pnma* centrosymmetric space group (improper ferroelectricity appears at a magnetic transition at ~28 K). Therefore, the perovskite YMnO$_3$-BiFeO$_3$ binary system seems a promising one to show a polar/nonpolar MPB.

The synthesis of the whole binary system was recently achieved for the first time by mechanosynthesis, and their structural characterization revealed the presence of two $Y_{1-x}Bi_xMn_{1-x}Fe_xO_3$ continuous solid solutions of orthorhombic and rhombohedral symmetries that coexisted in a broad range of compositions: x=0.5-0.9.[16] This scenario seems different to that of conventional MPBs as those of $Pb(Zr,Ti)O_3$ or $BiScO_3$-$PbTiO_3$,[18, 19] characterized by converging lattice distortions on approaching the boundary, presence of intermediate bridging phases and narrow compositional coexistence regions. However, wide phase coexistence has been anticipated by a first principles study of $BiFeO_3$-$BiCoO_3$, associated with two stable continuous solid solutions of rhombohedral and tetragonal symmetries with very close energies.[20] This case was referred to as "discontinuous" as compared with conventional "continuous" MPBs because lattice distortions increase on approaching the boundary. No bridging phases existed in this case.

Thermal stability was an issue, and phases decomposed under heating at moderate temperatures, making ceramic processing a challenge. This is required to accomplish a sound electrical characterization of the system, and to assess the existence of ferroelectricity for x < 1 (beyond $BiFeO_3$). We are aware of a single report of dielectric properties, concentrated in $0.8 \leq x \leq 1$, which used poorly characterized ceramic materials of unknown densification and microstructure.[25] However, or perhaps because of this, the presence of ferroelectricity was not addressed, and analysis exclusively concentrated on the magnetic properties.

In this work, we report the application of Spark Plasma Sintering to the ceramic processing of perovskite phases across the $YMnO_3$-$BiFeO_3$ binary system, starting from the nanocrystalline phases obtained by mechanosynthesis. This combination has been demonstrated a suitable procedure to obtain dense ceramics at low temperatures that helped in preventing the decomposition of thermally unstable phases.[26] Indeed, $Y_{1-x}Bi_xMn_{1-x}Fe_xO_3$ dense materials were obtained all across the system but for x = 0.2 and 0.3. This allowed characterizing the electrical properties of most phases, covering several orthorhombic phases, as well as all materials in the phase coexistence and rhombohedral regions. A special effort was put on assessing ferroelectricity, and on following the ferroelectric transition as x is decreased within the rhombohedral region. This required a combination of techniques encompassing electrical and mechanoelastic characterizations, along with temperature-dependent Raman spectroscopy.

**Experimental Procedure**

Twelve compositions across the system $YMnO_3$-$BiFeO_3$ were prepared by mechanosynthesis in a high-energy planetary mill: $Y_{1-x}Bi_xMn_{1-x}Fe_xO_3$ with x = 0, 0.1, 0.4, 0.5, 0.6, 0.7, 0.8, 0.9, 0.925, 0.95, 0.975 and 1. Stoichiometric mixtures of analytical grade commercial oxides $Bi_2O_3$, $Fe_2O_3$, $Y_2O_3$ and $Mn_2O_3$ were thoroughly ground by hand in an agate mortar, and placed in a Fritsch vario-planetary mill Pulverisette 4 (Fritsch, Idar-Oberstein, Germany) for mechanical treatment. Tungsten carbide vessels and balls were used. Additional details of the mechanical treatments, and of phase evolution during the process can be found in a previous work.[16] Perovskite mechanosynthesis was targeted, and phase evolution were controlled by X-ray powder diffraction, with a Bruker AXS D8 Advance diffractometer (Bruker, Karlsruhe, Germany). Patterns were collected between 12° and 60° (2θ), with 2θ increments of 0.05 and counting time of 0.2 s per step. The Cu-Kα doublet (λ = 1.5418 Å) was used in the experiments. Nanocrystalline perovskite phases were obtained for all compositions, which required tailoring milling time to complete the synthesis. Specific values are compiled in Table 1.

Ceramic processing of the different phases was performed by Spark Plasma Sintering. This advanced technique combines an uniaxial pressure and a pulsed direct electrical current (DC) under vacuum to obtain fast densification of powdered samples at reduced temperatures. This is enabled by the very high heating rates that cause densification to prevail over grain growth. A Dr Sinter lab. Jr Spark Plasma Sintering system, model 212Lx (Fuji, Tsurugashima, Japan) was used. In a typical experiment, a cylindrical graphite die with an inner diameter of 8 mm was filled with 0.6-0.9 g of the nanosized phases, depending on their density. A graphite paper was placed between the powder and the die to facilitate ceramic removal after the experiment. Uniaxial pressure was firstly increased up to 50 MPa in 1 min to improve the initial powder compaction. Then, the temperature was raised up to 100 ºC below the final target temperature (indicated in Table 1 for each sample) in 2 min, while pressure was maintained at 50 MPa. This was followed by further heating up to 50 ºC below the final temperature in 1 min, and loading up to 100 MPa in pressure. Final temperature was reached in an additional minute, and conditions were maintained for 2 min. Finally, cooling was carried out while maintaining pressure. The total process lasted about 14 min. The target temperatures for SPS were initially selected to be 50 ºC below the decomposition temperature of powdered phases in air. This temperature has been shown to decrease for $Y_{1-x}Bi_xMn_{1-x}Fe_xO_3$ orthorhombic phases with increasing x, while trend is reversed for rhombohedral ones.[13] However, secondary phases were present for some compositions after SPS with these initial conditions, so parameters were tailored until perovskite decomposition was prevented.

After SPS, the density of ceramics was measured by the Archimedes' method in distilled water. Phases in the sintered ceramics were also monitored monitored by X-ray powder diffraction, using a Bruker D8 diffractometer with a copper source operated at 1600 W (Bruker, Karlsruhe, Germany). A step size = 0.02 ° and exposure time = 0.5 s/step were selected. Ceramic microstructures were studied by field emission scanning electron microscopy on fractured surfaces after metallization with chromium using a Q150T-S sputtering system (Quorum Technologies, Laughton, United Kingdom). A Philips XL 30 S-FEG FE-SEM (Philips, Eindhoven, the Netherlands) apparatus with a 1-30 keV working range and a maximum resolution of 3.5 nm was used.

Electrical properties were characterized on ceramic discs thinned down to 0.5 mm thickness, on which Ag electrodes (8032 conductor paste, Du pont, Bristol U.K.) were painted and sintered at the same temperature used for SPS. Real and imaginary components of the permittivity were obtained from capacitance and loss tangent measurements, carried out with an E4980A precision LCR meter (Agilent Technologies, Santa Clara, CA/USA). This was done in two temperature ranges: (1) from 85 to 673 K, and (2) from RT to high temperature before phase decomposition. Measurements were dynamically carried out during heating/cooling cycles with ± 1.5°C min$^{-1}$ rate, at several frequencies between 100 Hz and 1 MHz. A cryostat Janis VPF 700 (Janis, Woburn, Massachusetts/USA), coupled to a temperature controller Lakeshore 331 (Lake Shore cryotronics, Columbus, Ohio/USA) and vacuum conditions were used in the low-temperature range, while a Seven Furnaces Ltd. 500 W power tubular furnace (Nabertherm, Lilienthal, Germany) coupled to a Eurotherm temperature controller (Schneider Electric, Rueil-Malmaison, France) with an error of ±0.5 ºC were used in the high-temperature range.

High-field electrical properties were also studied for rhombohedral phases ($x \geq 0.9$), whose polar crystal space group allows for ferroelectricity. Note that a fraction of orthorhombic phase coexists with the rhombohedral one for x=0.9, though it is minor. Experiments were carried out from RT down to 100 K, where minimal conduction contributions are expected as previously revealed by the temperature and frequency dependences of permittivity. Low frequency (0.1 Hz) voltage sine waves with amplitude up to 10 kV were applied by the combination of a synthesizer/function generator HP 3325B (Agilent Technologies, Santa Clara, CA/USA) and a high voltage amplifier TREK model 10/40 (TREK, Lockport, New York/USA), and charge was measured with a homebuilt charge-to-voltage converter and software for loop acquisition and analysis.

As anticipated, conduction mostly dominated the high-temperature electrical response of all phases, hindering the identification of possible dielectric anomalies associated with the ferroelectric transitions in the rhombohedral region. Alternatively, and taking advantage of the electromechanical coupling, one can look for the related elastic anomaly in the Young´s modulus, investigated by dynamical mechanical analysis (DMA). This technique has been shown to be very suitable for studying phase transitions and the dynamics of domain walls in ferroelectrics. [27–29] The low-frequency Young's modulus and mechanical losses were measured as a function of temperature by DMA in three-point bending configuration with a Perkin Elmer DMA7 apparatus (Perkin Elmer, Waltham, Massachusetts/USA). A stress sine wave of 2.6 MPa amplitude, superimposed on static stress of 3.4 MPa, was applied to ceramic bars of 12 x 2 x 0.5 mm$^3$ dimensions. Measurements were dynamically accomplished at a single frequency of 33 Hz during a heating/cooling cycle: from RT to high temperature before perovskite decomposition, and back to RT with ± 2 K min$^{−1}$ rates.

Additionally, temperature-dependent Raman measurements were performed for selected compositions to demonstrate the relation of the dielectric and elastic anomalies with the ferroelectric transition. A LabRAM (Horiba Jobin Yvon, Villeneuve d'Ascq, France) spectrometer with a 532 nm laser excitation were used. The laser beam spot had a diameter of ~1 μm on the specimen surface. The samples were inserted in a Linkam MS600 stage (Linkam, Tadworth, UK) for temperature-dependent measurements. Spectra were collected in backscattering geometry with an 1800 gr/mm grating and a Peltier-cooled charged coupled device (CCD) camera, allowing a spectral resolution of at least 1.5 ± 0.1 cm$^{-1}$/pixel for the investigated range. The measured spectra were deconvoluted with a sum of Gaussian– Lorentzian peak functions in a commercial software environment (Labspec 4.02; Horiba Jobin Yvon).

Finally, the magnetic properties of rhombohedral phases were characterized by a Quantum Design MPMS-XL5 SQUID magnetometer (Quantum Design, San Diego, California/USA). Magnetization was measured between 2 and 400 K after zero-field cooling (ZFC) and during subsequent field cooling (FC) under an applied field of 100 Oe, and as a function of magnetic field at room temperature.

**Results and Discussion**

*Spark Plasma Sintering*

In this work, the ceramic processing of the mechanosynthesized phases across the perovskite YMnO$_3$-BiFeO$_3$ binary system was studied by Spark Plasma Sintering. Formation of secondary phases took place during the SPS when processing temperatures were maintained at 50 °C below the maximum T above which perovskite phases decompose in air, as determined previously using conventional furnaces. This might be due to the reducing conditions created during SPS by the presence of graphite, and favoured by the dynamic vacuum during the process. Temperatures were then decreased until second phases were not observed. This resulted in densifications below 90% (and thus open porosity) at 100 MPa for a number of compositions, so pressure was increased in these cases aiming to enhance density. This required substituting graphite dies by tungsten carbide ones. Final SPS conditions and resulting densification values are summarized in Table 1. Note that parameters are not given for Y$_{1-x}$Bi$_x$Mn$_{1-x}$Fe$_x$O$_3$ with x=0.2 and 0.3, for which materials dense enough (above 90%) to be suitable for electrical characterization were not obtained. Nonetheless, high-quality ceramic materials were processed for several orthorhombic phases, including x = 0 and 0.1, as well as for all compositions across the phase coexistence and rhombohedral regions. As far as we know, this is the first work that shows how to obtain dense ceramics of orthorhombic YMnO$_3$ from a metastable orthothombic YMnO$_3$ powders. X-ray diffraction (XRD) patterns of all sintered ceramics are shown in Figure 1. A distinctive evolution of the perovskite lattice, consistent with that described for powdered phases,[16] is observed from an orthorhombic *Pnma* symmetry (x < 0.5) to a rhombohedral *R*3*c* symmetry (x > 0.9), showing the coexistence of both symmetries between x = 0.5 and x = 0.9. Gradual shift of all diffraction peaks with x reflect the continuous evolution of the lattice parameters across both orthorhombic and rhombohedral solid solutions, and strongly supports the presence of a MPB.

Table 1: Required reaction time in mechanosynthesis (t), maximum temperature phases can withstand before decomposition (T Max.), thermal treatment of powdered phases (TT), Spark Plasma Sintering conditions (temperature, T, and pressure, P) and densification values (D) of the obtained ceramic materials. Latter parameter refers to bulk density expressed as percentage of the crystallographic density, obtained from XRD data.

|         | t (h)[16] | T Max. (°C)[16] | TT  | T (°C) | P (MPa) | D (%) |
|---------|-----------|-----------------|-----|--------|---------|-------|
| BiFeO$_3$ | 14        | 750             |     |        |         | > 99  |
| 0.975   | 16        | 825             |     |        |         | > 99  |
| 0.95    | 17        | 825             | Yes | 600    | 100     | > 99  |
| 0.925   | 18        | 825             |     |        |         | > 99  |
| 0.9     | 20        | 750             |     |        |         | >98   |
|         |           |                 |     | 600    | 100     | > 99  |
| 0.8     | 11        | 700             |     | 600    | 100     | 99    |
| 0.7     | 9         | 600             |     | 550    | 100     | 92    |
| 0.6     | 7         | 550             |     | 500    | 200     | 95    |
| 0.5     | 5         | 550             | No  | 500    | 250     | 90    |
| 0.4     | 3         | 550             |     | 475    | 350     | 92    |
| 0.3     | 4         | 550             |     | --     | --      | --    |
| 0.2     | 5         | 600             |     | --     | --      | --    |
| 0.1     | 7         | 700             |     | 700    | 300     | 91    |
| YMnO$_3$ | 8         | 900             |     | 700    | 300     | 93    |

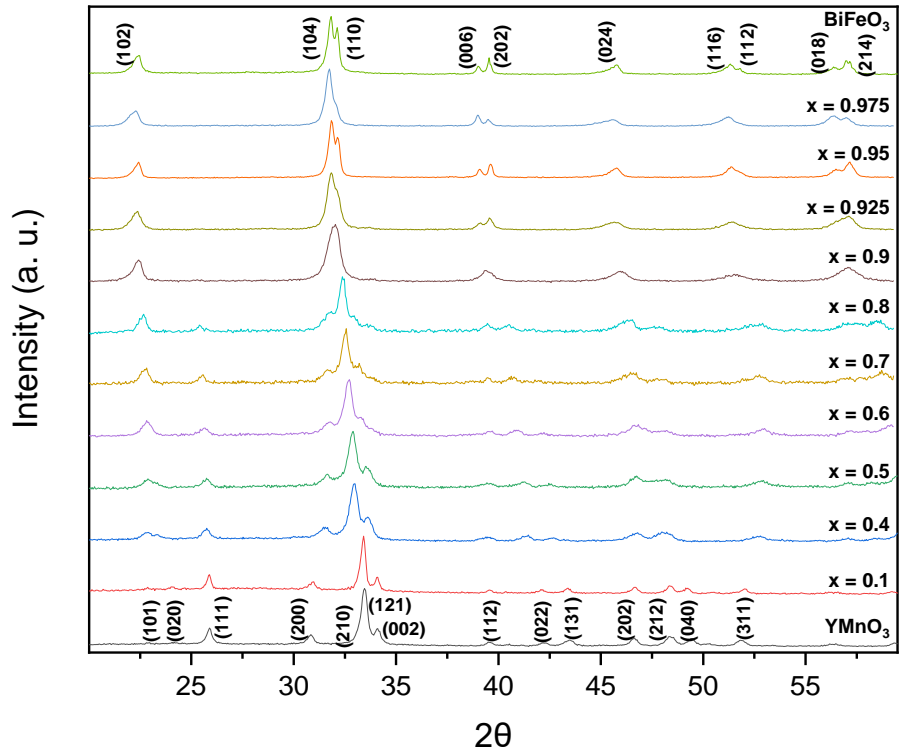

Figure 1: XRD patterns for the $Y_{1-x}Bi_xMn_{1-x}Fe_xO_3$ ceramics processed by SPS. Miller indices corresponding to the *Pnma* and *R3c* crystal structures are indicated for $YMnO_3$ and $BiFeO_3$, respectively.

Ceramic microstructures could not be studied in polished surfaces, because neither thermal etching plus quenching, nor chemical etching succeeded in revealing grain structure as illustrated in Fig. 2a for x = 0.9. A Backscattered electron image is provided in Fig. 2b to indicate the high chemical homogeinity related to mechanosynthesis. Microstructures were then studied on fractured samples, and images of selected compositions are also given in figure 2. Intergranular fracture modes were obtained for all cases, and no significant differences in grain size resulted. Grain sizes were evaluated with the average grain intercept method in SEM micrographs. Values ranging between 30 and 50 nm were found, deep into the nanoscale. Note that nanostructuring was also obtained for x =0.1 (Figure 2d), sintered at a comparatively high temperature of 700 °C. This is most probably an effect of the very high pressure (300 MPa) required for densifying this orthorhombic phase, which strongly inhibits grain growth.[30]

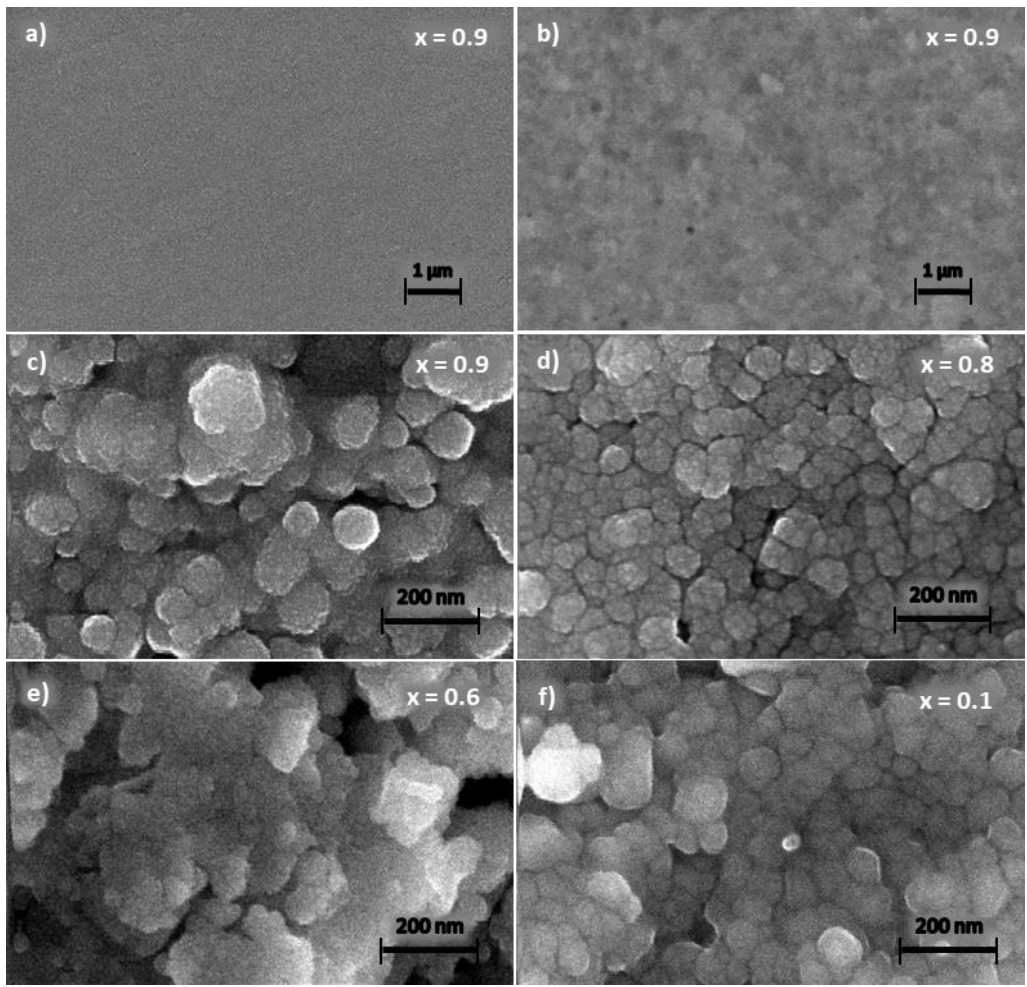

Figure 2: SEM micrographs for the $Y_{1-x}Bi_xMn_{1-x}Fe_xO_3$ ceramics: of polished surfaces (a) x = 0.9 secondary electron image, (b) x = 0.9 backscattered electron image; and of fresh fractures (c) x = 0.9, (d) x = 0.8, (e) x = 0.6, (f) x = 0.1

Ceramic nanostructuring of ferroelectrics is known to cause a strong depletion and broadening of the dielectric anomaly associated with the transition, until its disappearance below a certain threshold. [31, 32] Therefore, the processing of ceramics with coarsened microstructure was studied in the rhombohedral region, with the objective of uncovering eventual dielectric anomalies associated with the ferroelectric transition. This was done by SPS of phases that had been thermally treated at 750 °C (x = 0.9, 1) and 825 °C (x = 0.925, 0.95, 0.975), which resulted in particle sizes in the range of 0.5 - 0.7 µm. Indeed, dense ceramics with an average grain size between 0.6 and 0.8 µm were obtained, as illustrated for x = 0.95 in Figure 3a. It is remarkable that densification of the submicron-sized particles is attained by SPS with hardly any grain growth. A comparison of the densification kinetics of the mechanosynthesized (MS) and the thermally treated (TT) phases is also given in Fig. 3b. Temperature, pressure and shrinkage curves are all displayed. Note the faster densification rate of the nanocrystalline powder, as compared with the submicron-size one. Despite this difference in kinetics, very high densification was achieved with both powders using the same SPS parameters in a total processing time as short as 14 min.

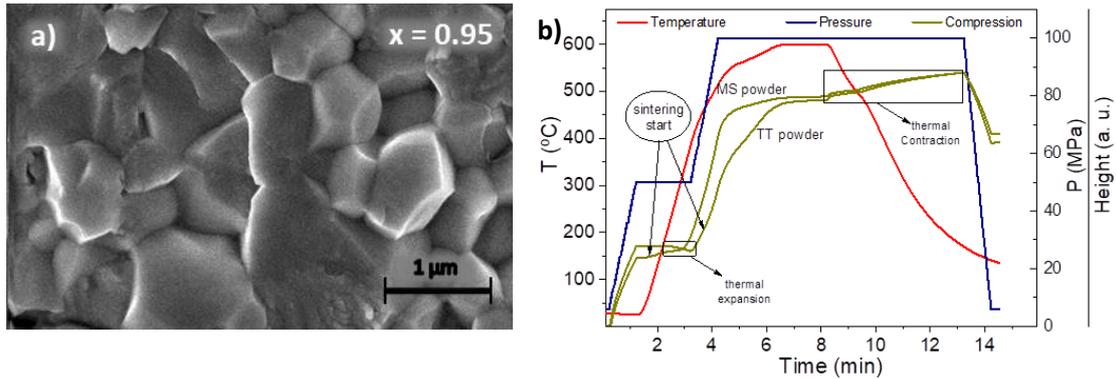

Figure 3 (a) SEM micrograph of a ceramic sample with x=0.95 and coarsened microstructure. (b) Comparison between in-situ shrinkage curves during the SPS of $Y_{1-x}Bi_xMn_{1-x}Fe_xO_3$ as mechanosynthesized (MS) and thermally treated (TT) phases with x = 0.9.

*Electrical characterization*

Measurements of the dielectric permittivity and losses as a function of temperature and frequency revealed the presence of thermally activated conduction, already significant at room temperature for all compositions studied. This is illustrated in Figure 4, where the temperature dependences of the relative permittivity and loss tangent are given for $Y_{1-x}Bi_xMn_{1-x}Fe_xO_3$ ceramic samples with x=0.4, 0.7 and 0.95 at several frequencies, as examples. These specific compositions correspond to the orthorhombic, phase coexistence and rhombohedral regions, respectively. Note the high dispersion with frequency for the three cases, directly reflecting conduction. Dielectric losses are proportional to the DC conductivity and scale with the reciprocal frequency. An evolution of the onset of dispersion is found so that the permittivity at 100 Hz starts deviating from that at 1 kHz below 100 K for x=0.4, and at ~125 and ~200 K for x=0.7 and 0.95, respectively. This indicates a decreasing conductivity when x is increased. Low impedance was an issue at high temperature, and permittivity could only be obtained until a certain temperature.

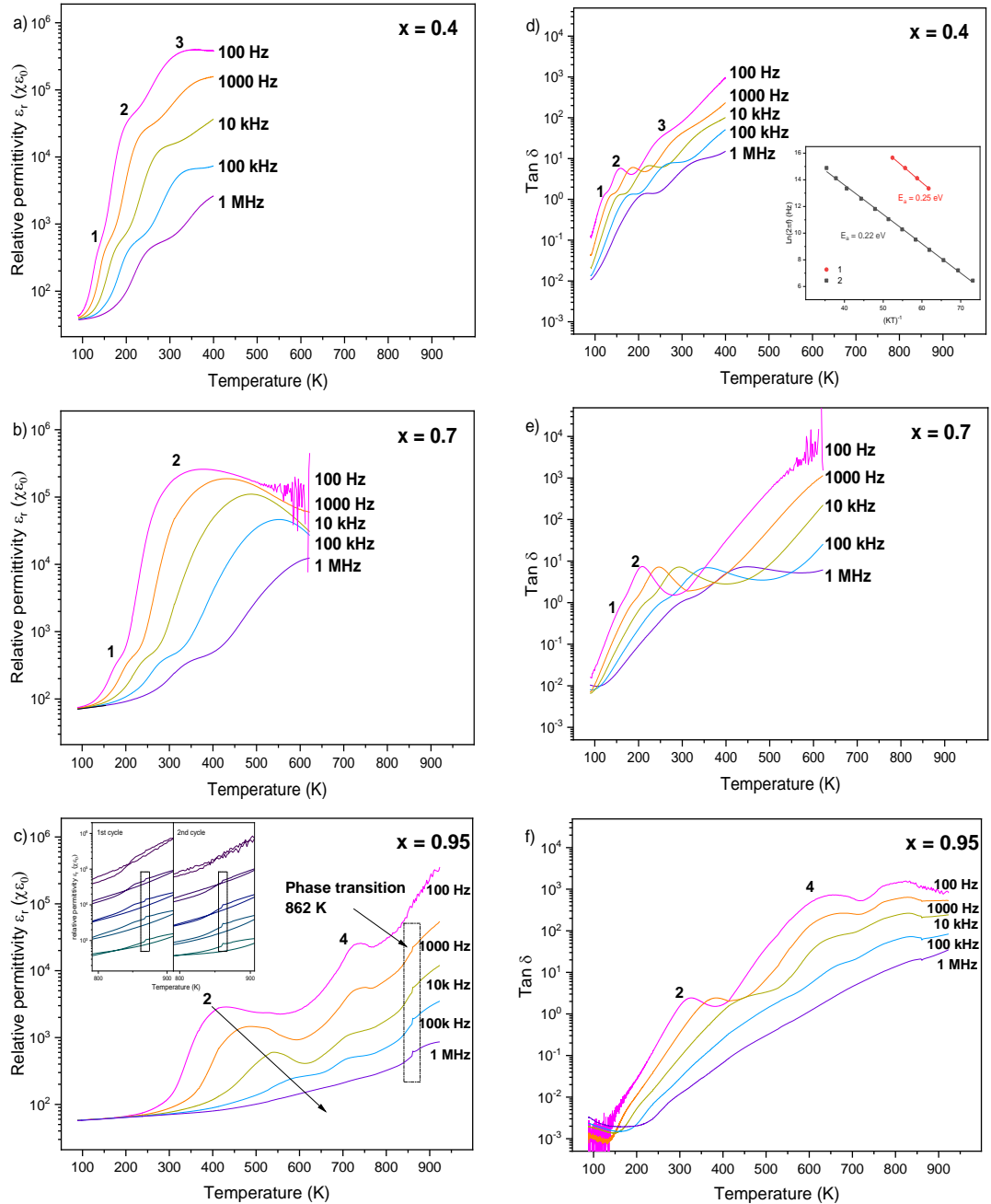

Figure 4: Temperature dependences of the real permittivity (a-c) and loss tangent (d-f) for the $Y_{1-x}Bi_xMn_{1-x}Fe_xO_3$ ceramics with x=0.4 (a,d), x=0.7 (b,e) and x=0.95 (c,f), measured during heating at several frequencies (0.1–1000 kHz). Inset in (c) shows the thermal hysteresis around a high temperature anomaly that could be associated with the ferroelectric transition, while that in (d) shows Arrhenius analysis of the two dielectric relaxations labelled as 1 and 2.

In the case of BiFeO$_3$, conductivity has been related to the electron hopping between Fe$^{2+}$/Fe$^{3+}$ cations, Fe$^{2+}$ resulting from the formation of oxygen vacancies.[33] However, a recent report demonstrated the presence of Fe$^{4+}$ and A-site vacancies at ferroelectric domain walls, and proposed a mechanism of confined Fe$^{3+}$/Fe$^{4+}$ hopping along the walls.[34] High conductivity in BiMnO$_3$-PbTiO$_3$ has been associated with the presence of an oxygen hyperstoichiometry and thus, of Mn$^{4+}$. A mechanism of electron hopping between Mn$^{3+}$ and Mn$^{4+}$ was proposed in this case.[15, 35] Both mechanisms might be active across the YMnO$_3$-BiFeO$_3$ system, and alternatively dominate response.

Dielectric characterization also revealed the presence of several successive dielectric relaxations (labelled as 1, 2 and 3 in the figure). See firstly the dielectric response of the material with x=0.4: There are two successive step-like increases of permittivity on heating from 85 K, whose height and position decrease and shift to high temperatures, respectively, with frequency. Steps in the relative permittivity are accompanied with maxima in the loss tangent (also marked as 1 and 2). This is characteristic of Maxwell-Wagner (M-W) type relaxations, associated with the blocking and accumulation of free charge carriers at interfaces, usually grain boundaries in the case of ceramic materials.[36, 37] The set of maxima in loss tangent (T$_m$) allows relaxation frequency to be obtained as a function of temperature. This was done for the two successive M-W relaxations (effects 1 and 2), and results are depicted in the inset of Fig. 4d. Good fits to an Arrhenius law, given by Eq. (1):

$$f = f_0 \exp\left(-E_a/k_B T\right) \qquad (1)$$

resulted, which provided the activation energy $E_a$ for both processes. Similar activation energy of 0.22-0.25 eV was obtained, which strongly suggests that the same charge carriers are involved (and successively blocked at two boundaries).

The same two M-W effects are found for the material with x = 0.7, while only relaxation 2 is observed for x=0.95. There is also a distinctive shift of the M-W relaxations towards higher temperature with x that is consistent with the varying concentrations of carriers, but also with increasing grain size. SPS temperatures were 475 and 550 ºC for x=0.4 and 0.7, respectively. The material with x= 0.95 is a coarsened ceramic, processed from powders that are thermally treated prior to SPS. Relaxation frequencies as a function of temperature could only be obtained for all compositions in the case of effect 2, and are given in Fig. 5a. Good fits to the Arrhenius law also resulted, and the obtained activation energies for this relaxation are presented in Fig. 5b as a function of composition. Only exception was x = 0.975, for which reliable $T_m$ values could not be extracted. A distinctive trend in $E_a$ with composition is revealed. An activation energy of $\approx 0.31$ eV was found for $YMnO_3$ (x=0), which agreed with the activation energy value for the DC conductivity in the same material, as expected for a M-W type relaxation. Recall that this activation energy corresponds to effect 2, which is most probably associated with the block of charge carriers at grain boundaries. DC conductivity was calculated by complex impedance analysis of conductance data as a function of frequency at different temperatures,[32] as shown in Fig. 5c. This number is also in the range of activation energies of other related oxides with $Mn^{3+}/Mn^{4+}$ hopping conduction [15, 21, 35, 36]. Low activation energies between 0.22 and 0.35 eV were obtained between x=0.1 and 0.7 with no clear trend. However, from x=0.8 activation energy continuously increased, and reached a maximum of $\approx 0.45$ eV for x = 0.95. Once again, this value agrees well with the activation energy of the DC conductivity for high-x compositions (see Fig. 5c), and with values for related oxides with $Fe^{2+}/Fe^{3+}$ hopping conduction.[21] This seems to confirm the presence of the two different charge carriers in the system, which alternatively dominate conduction as composition is changed. It is remarkable the low activation energy obtained for x=0.4 and 0.5, because it is below the corresponding $E_a$ for the DC conductivity in these cases. An explanation cannot be given at this point, though errors in the determination of the set of $T_m$, associated with the overlapping of the successive dielectric processes, could have led to underestimating activation energy.

Beyond these two relaxations, a third one takes place for x=0.4 (labelled as 3) that rather resembles a Debye-type relaxation than a M-W one. This third event is not observed for the material with x=0.7, though the decrease of permittivity above the second M-W relaxation could indicate its presence but overlapped with the M-W effect. There is a fourth dielectric anomaly (labelled as 4 in fig. 4c and f) for the x=0.95 material at high temperature, irreversible (not observed on subsequent cooling), and at a temperature that does not change with frequency. A similar effect in the $BiFeO_3$-$BiMnO_3$-$PbTiO_3$ ternary system has been associated with the sudden release of trapped charges.[23]

A sound dielectric permittivity can only be obtained at low temperature, before conduction losses become significant. This was done at 100 K, and its variation across the system is given in figure 6 at 1 kHz and 1 MHz. Neligible or little dispersion was found at this temperature, which rules out possible extrinsic contributions related to accumulation of charge carriers at grain boundaries, or to active dipolar defects and low temperature Debye relaxations. Relative permittivity values of $\approx$ 15 and $\approx$ 50 were found for $YMnO_3$ (x=0) and $BiFeO_3$ (x=1). These values compare well with literature ones for orthorhombic $YMnO_3$ ($\approx$ 20)[11] and rhombohedral $BiFeO_3$ ($\approx$ 40).[37] Remarkably, permittivity increases linearly with x in the orthorhombic region, is further enhanced in the phase coexistence region, and decreases after entering the rhombohedral one. Therefore, there is a distinctive maximum for x=0.8-0.9 that coincides with phase coexistence. Maxima in permittivity are well known to take place at ferroelectric and multiferroic MPBs in perovskite binary systems [37–39], and have been shown here to also occur for $YMnO_3$-$BiFeO_3$ at the polar/non-polar MPB region. This suggests that the polarization extension mechanism, giving place to property enhancement,[40] might be active in $YMnO_3$-$BiFeO_3$.

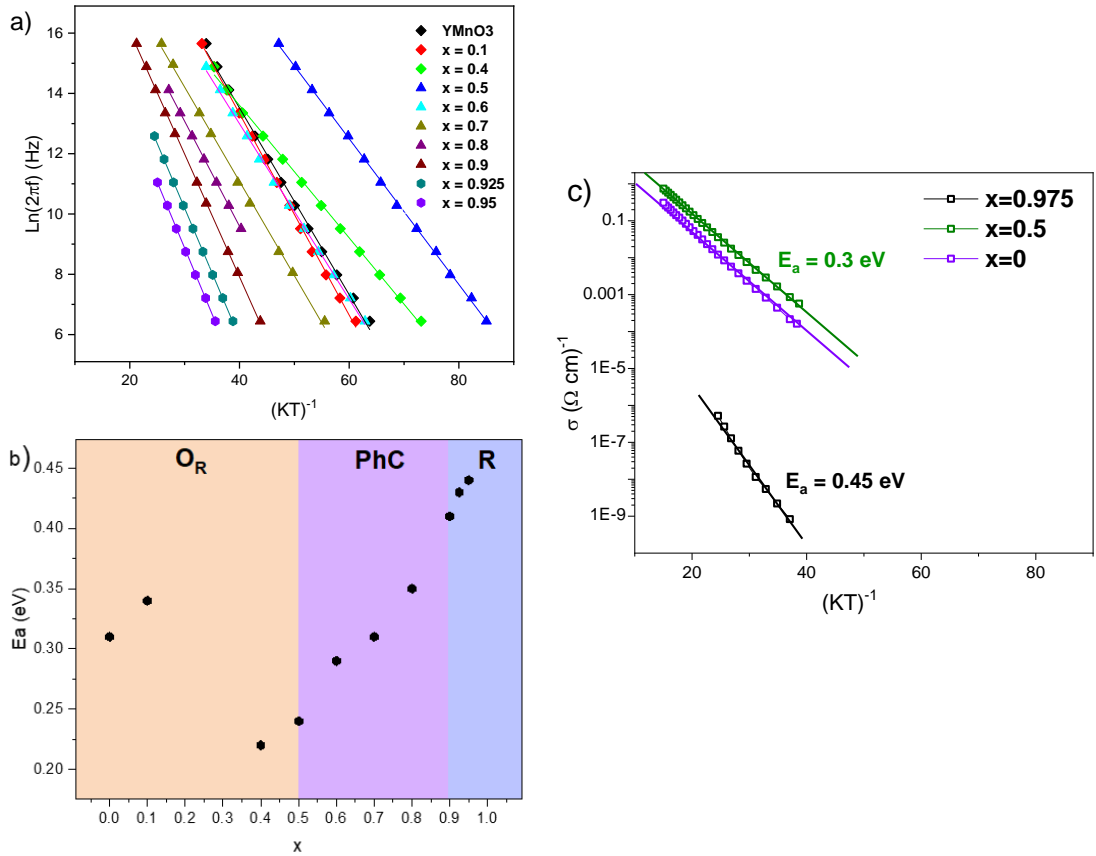

Figure 5: (a) Arrhenius plots corresponding to the second Maxwell-Wagner relaxation for the $Y_{1-x}Bi_xMn_{1-x}Fe_xO_3$ ceramics, (b) obtained activation energies as a function of composition across the binary system (orthorhombic ($O_R$), phase coexistence (PhC) and rhombohedral (R) regions are indicated), and (c) Arrhenius analysis of dc conductivity for selected compositions.

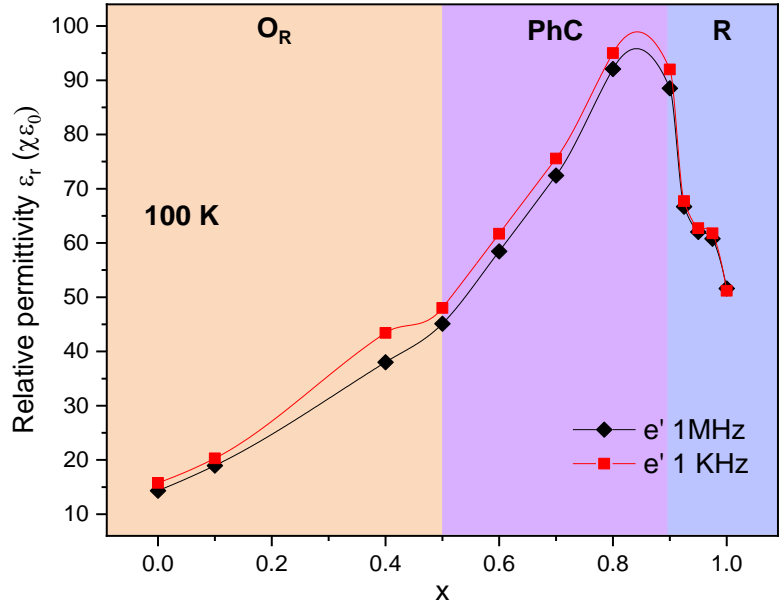

Figure 6: Low temperature relative permittivity for the $Y_{1-x}Bi_xMn_{1-x}Fe_xO_3$ ceramics at two frequencies. Orthorhombic ($O_R$), phase coexistence (PhC) and rhombohedral (R) regions are indicated.

Extensive conduction at room temperature prevented the high field response to be characterized. Low-temperature measurements were then carried out for compositions x = 0.9, 0.925, 0.95 and 0.975, that were thought to be rhombohedral *R*3*c*, searching for ferroelectric hysteresis loops. Note that the orthorhombic *Pnma* is centrosymmetric and thus, it does not allows for ferroelectricity. Following previous work on other $BiFeO_3$-based systems, ceramic samples were quenched from temperatures above the transition temperature to release domain walls, usually pinned in these systems.[37] However, no ferroelectric switching was found before the dielectric breakdown, even with applied fields as high as 12 kV mm$^{-1}$. This is illustrated in Fig. 7a for x = 0.975 at 175 K, as an example. Note the field-dependent effective permittivity, typical behaviour of ferroelectric materials in the subcoercive regime. Actually, permittivity linearly increased with field amplitude for all measured ceramics, behaviour that can be related to a Rayleigh process.[41] This trend is shown in Fig. 7b, and proves the ferroelectric nature of all $Y_{1-x}Bi_xMn_{1-x}Fe_xO_3$ rhombohedral phases with x ≥0.9. Rayleigh behaviour results from the reversible movement of ferroelectric domain walls across a field of defects, with which they interact. Like dielectric permittivity at 100 K, the high field effective permittivity also increases as x decreases across the rhombohedral region, up to a value of ~225 for x=0.9 (under a driving electric field of 9 kV mm$^{-1}$).

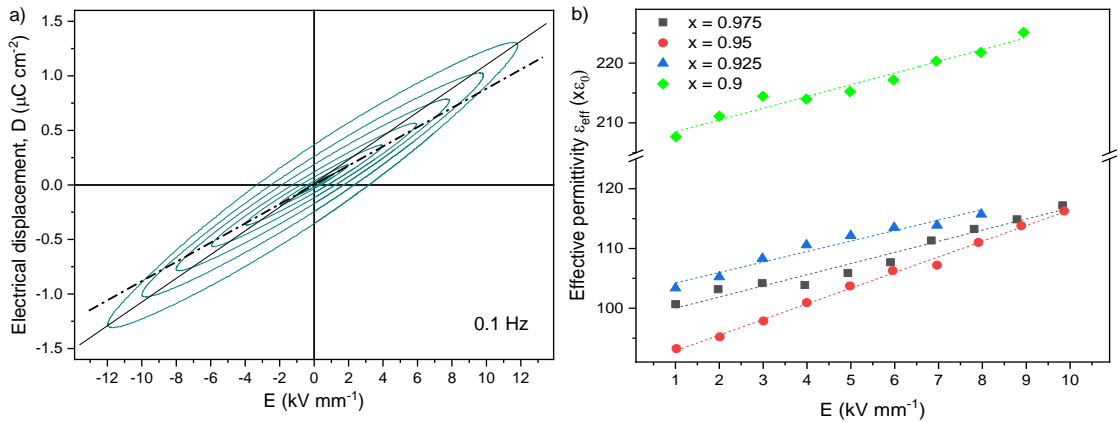

Figure 7: (a) High field electrical response for a Y$_{1-x}$Bi$_x$Mn$_{1-x}$Fe$_x$O$_3$ ceramic with x=0.975 at 175 K, and (b) effective permittivity at increasing driving electric fields for Y$_{1-x}$Bi$_x$Mn$_{1-x}$Fe$_x$O$_3$ materials in the rhombohedral region.

All rhombohedral phases are thus ferroelectric. An obvious question then is how the Curie temperature evolves with x, which is known to be ~ 825 ºC in the case of BiFeO$_3$ (x=1).[5] Indeed, $T_c$s of 594 ºC and 509 ºC have been reported on heating and cooling, respectively, for x=0.95 from DTA measurements on thermally treated powdered samples. On the contrary, the thermal events associated with the transition were not observed for x=0.925.[16] Analogous measurements were carried out in this work for x=0.975 and 1, and results are shown in Fig. 8a for powdered materials treated at 825 ºC. The curve for x=0.95 from ref. [16] is included for comparison. Relating ceramics, the temperature dependence of permittivity shows a distinctive anomaly at the ferroelectric transition of BiFeO$_3$,[37] and a similar effect was found at 862 K for x = 0.95 during the heating run (see inset in Fig. 4c). This value agrees with that obtained by DTA on powdered samples, but the anomaly in permittivity was not observed on subsequent cooling, as it would be expected for a reversible phase transition (even if it consistently reappears in successive heatings). Likewise, anomalies were not observed for x=0.9, 0.925 and 0.975. Issue is likely conductivity and dielectric relaxations at high-temperature that mask the anomalies associated with the transition. An alternative technique was then needed to characterize the ferroelectric transition in the ceramic materials.

*Mechanoelastic characterization and Raman measurements*

The ferroelectric transition in the rhombohedral phases was further investigated by dynamical mechanical analysis. Results of the low frequency Young's modulus for $Y_{1-x}Bi_xMn_{1-x}Fe_xO_3$ ceramics with x = 0.90, x = 0.925 and x = 0.95 are shown in Fig. 8b. The three materials presented a reversible elastic anomaly consisting of a step-like hardening on heating with significant thermal hysteresis, at a temperature that increased with x. The anomaly is much sharper for x=0.925 and 0.95 than for x=0.9, for which a temperature interval 180-300°C can only be defined. Note that a minor fraction of the orthorhombic phase was already present for this composition. On the contrary, anomalies took place at $\approx$ 400 and 595 °C on heating (330 and 520 °C on cooling) for x=0.925 and 0.95, respectively. These values for x = 0.95 are in a good agreement with the high-temperature dielectric anomaly on heating and with transition temperatures obtained by DTA analysis of powdered samples (Fig. 4c and Fig. 8a, respectively). It is then justified to relate the reversible elastic anomalies to the ferroelectric phase transition and to define the transition temperature $T_c$ as the one at which they occur. The obtained temperatures are collected in Fig. 8c, where values determined by dynamical mechanical analysis on ceramics are shown, along with those provided by DTA of powdered samples.

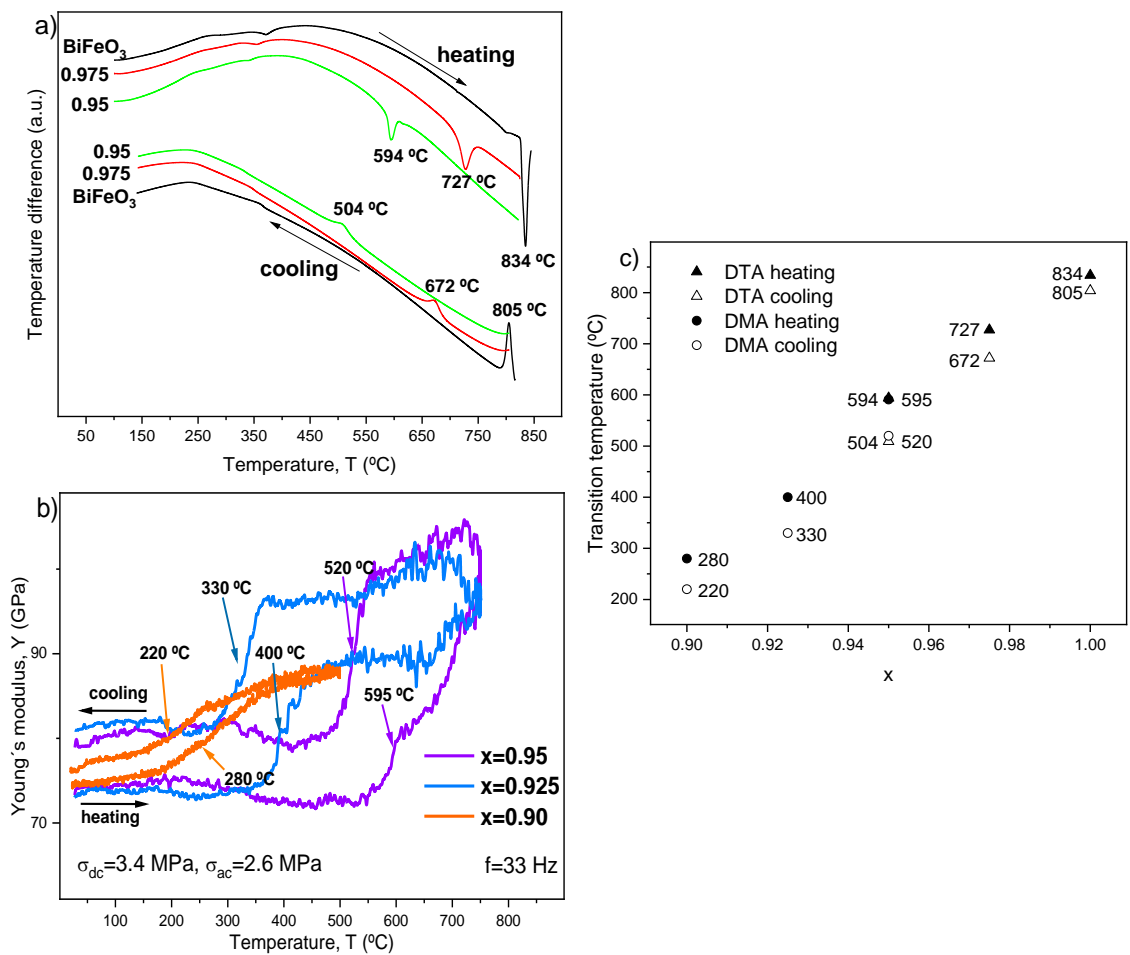

Figure 8: Determination of the ferroelectric transition temperature in the rhombohedral region, (a) thermal anomalies for crystalline powdered samples by DTA, (b) elastic anomalies for ceramic materials by DMA, and (c) resulting compositional dependence of the transition temperature.

The transition temperatures were confirmed by Raman spectroscopy for x=0.9 and 0.925 (the determined transition temperature for x=0.95 was at the limit of the operation range of the equipment, 600 °C, and thus was not possible to determine). Raman spectra at increasing temperatures are shown in Fig. 9. Note firstly the room temperature ones. According to bibliography,[16, 42] two transversal optic phonon modes [$E_g(TO)^1$ = 137 cm$^{-1}$; $A_{1g}(TO)^1$ = 143 cm$^{-1}$] and one longitudinal optic phonon modes [$E_g(LO)^2$] = 174 cm$^{-1}$] can be assigned to the *R*3*c* symmetry in BiFeO$_3$ and rhombohedral YMnO$_3$-BiFeO$_3$ (marked in the figure). These Raman modes correspond predominantly to Bi motions, according to literature.[43] *R*3*c* modes progressively disappear on heating, so that they cannot be observed from ~ 150 and 280 °C for x=0.9 and 0.925, respectively. On subsequent cooling, rhombohedral modes reappear. This evolution can be attributed to a structural transformation from ferroelectric *R*3*c* to paraelectric *Pnma* symmetry and thus, to the ferroelectric transition. Note that A-site and B-site substitutions in ABO$_3$ perovskite phases produce short-range lattice distortion,[44] and result in broad Raman modes. This makes the determination of a transition temperature difficult. Nonetheless, apparent temperatures are slightly lower than those of the elastic anomalies. This is most probably associated with ceramic materials being black in colour, so that strong light absorption takes place under the incident laser beam that causes local heating, and actual temperatures to be higher than nominal ones.[45] Mechanical anomalies then can be unambiguously associated with the ferroelectric phase transition that is of the R3c → Pnma type in this case. Note that additional, irreversible hardening is triggered above 650 °C for both x=0.925 and x=0.95 compositions. Actually, this effect was also observed for x=0.975 and resulted in massive plastic deformation that prevented locating the elastic anomaly (note that the thermal effect in DTA was observed at 727 °C for a powdered sample). The origin of this effect is under investigation, but the high-temperature orthorhombic phase of $BiFeO_3$ has been suggested to be ferroelastic so that additional hardening could indicate the coercive stress being surpassed at high temperature.

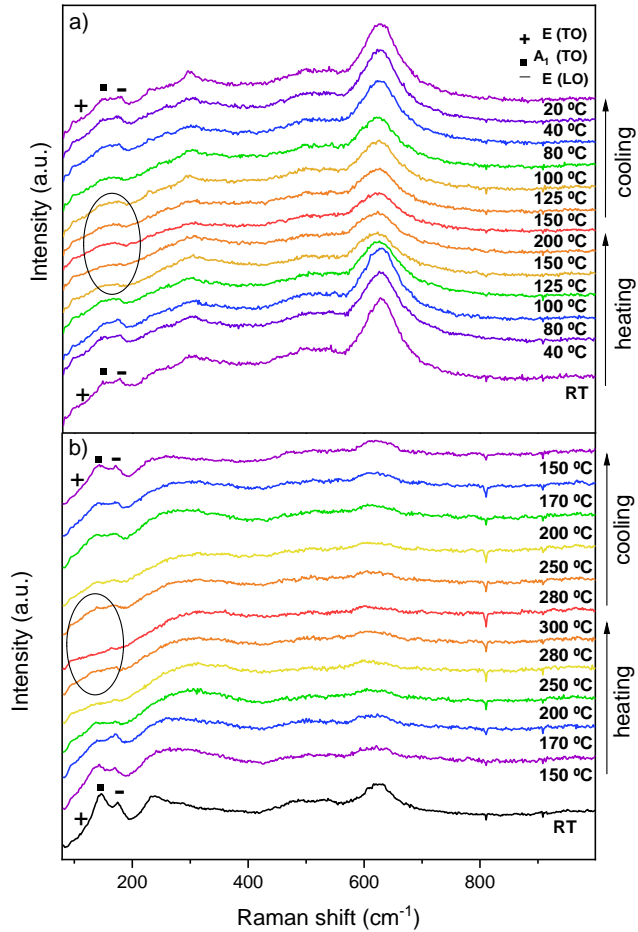

Figure 9: Raman spectra as a function of temperature for $Y_{1-x}Bi_xMn_{1-x}Fe_xO_3$ ceramics with (a) x=0.9, and (b) x=0.925 (Rhombohedral $E_g(TO)$, $A_{1g}(TO)$ and $E_g(LO)$ optic phonon modes are labelled as **+**, ▪ and **-** , respectively).

The presence of a high temperature polymorphic phase transition (PPT) to a orthorhombic structure for all rhombohedral phases from x=0.9 up to 1, and its shift towards low temperature as x is decreased raise the possibility that phase coexistence does not indicate an actual MPB, but it is just the consequence of the PPT temperature dropping below room temperature. Nonetheless, no hint of the ferroelectric transition was found for x=0.8 down to 77 K. Besides, a narrow width of the phase coexistence would be expected in this case from the high chemical homogeneity of the samples. Large width strongly suggests the presence of a "discontinuous" MPB as that anticipated for $BFeO_3$-$BiCoO_3$ in a first principles study.[20] *Magnetic measurements*

Rhombohedral phases are all, as *R3c* $BiFeO_3$ (x=1) is, room temperature ferroelectrics and then, they would be multiferroics if the antiferromagnetism of $BiFeO_3$ is maintained. The presence of magnetic ferroic order was investigated by temperature-dependent magnetization measurements. Zero field-cooling (ZFC)/field-cooling (FC) curves are shown in Fig. 10 for all compositions between x=0.9 and 1 under a magnetic field of 100 Oe. Qualitative differences were not found, and magnetic behaviour was analogous to that of $BiFeO_3$, known to be antiferromagnetic (AFM) with a $T_N$ of 370 ºC,[46] for all cases. Main effect of decreasing x is a distinctive increase of magnetization along with irreversibility, which suggests the appearance of spin canting and weak ferromagnetism, as observed for other modifications of $BiFeO_3$.[47–49] This magnetization and irreversibility were maximized for x=0.925 and decreased for x=0.9. Results shown here are consistent with those recently reported for phases obtained by conventional solid state synthesis.[50] In that work, room temperature M-H loops were measured for $Y_{1-x}Bi_xMn_{1-x}Fe_xO_3$ with 0.9 ≤ x ≤ 1, and a maximum magnetization was also found for x=0.925. This behaviour was related with the destruction of the spin cycloid of BiFeO$_3$, and the structural evolution of the $R3c$ structure with composition, through variations in the $Fe^{3+}$-$O^{2-}$-$Fe^{3+}$ bond angles and their effect in the superexchange interactions. Decrease for x=0.9 was associated with the presence of a minor fraction of the orthorhombic phase, which would then not be AFM.

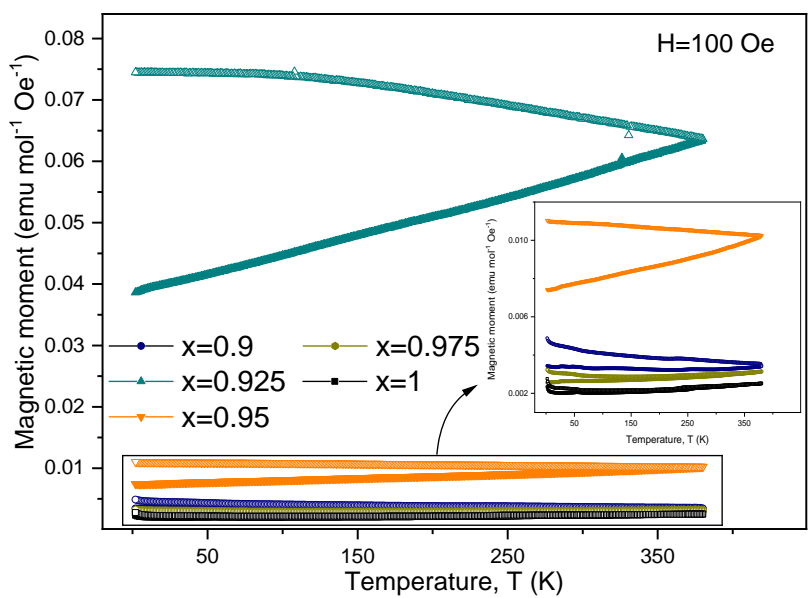

Fig. 10. ZFC (closed symbols) and FC (open symbols) temperature dependence magnetization curves for $Y_{1-x}Bi_xMn_{1-x}Fe_xO_3$ phases in the rhombohedral region. Inset shows the figure enlargement of magnetic moment values from 0.0015 to 0.012 emu mol$^{-1}$ Oe$^{-1}$.

**Conclusions**

In conclusion, high-quality ceramic materials, suitable for electrical characterization, could be processed for most phases of the perovskite YMnO$_3$-BiFeO$_3$ binary system by spark plasma sintering of nanocrystalline powders obtained by mechanosynthesis. These included $Y_{1-x}Bi_xMn_{1-x}Fe_xO_3$ orthorhombic phases like YMnO$_3$ (x=0) and x=0.1, as well as all phases with x ≥ 0.4, covering the phase coexistence and rhombohedral regions. Low temperatures, short times and high pressures were required to avoid phase decomposition, which resulted in ceramic nanostructuring. Nonetheless, submicron grain-sized materials of the rhombohedral compositions could be obtained without modifying SPS conditions, by starting from thermally treated powders. Dielectric characterizations revealed thermally activated conduction, already significant at room temperature that mostly controlled the electrical response. However, sound dielectric permittivities were obtained at 100 K, and a distinctive dependence with composition was shown. Namely, relative permittivity linearly increased with x in the orthorhombic region, it was further enhanced across phase coexistence, and decreased in the rhombohedral one, so that maximum values were obtained for x = 0.8-0.9. This might be associated with the presence of a polar-nonpolar MPB region in the system, at which polarizability enhancement would take place by the polarization extension mechanism.

Low temperature, high–field electrical measurements for rhombohedral compositions revealed a field-dependent effective permittivity that followed a Rayleigh behaviour, characteristic of ferroelectric materials. The ferroelectric transition was located by mechanoelastic measurements on the ceramic samples and confirmed with temperature-dependent Raman spectroscopy, and by differential thermal analysis on powdered materials. A decrease with x was found, so that $T_C$ decreased from ~ 825 ºC for x=1 ($BiFeO_3$) down to ~ 250 ºC for x = 0.9, well above room temperature. Besides, magnetic characterization revealed canted antiferromagnetism for the same compositions and then, multiferroism.

**Acknowledgements**

This work was supported by Spanish MICINNU through the MAT2017-88788-R project. José Ángel Quintana-Cilleruelo is also thankful for the financial support by the Spanish MINECO (BES-2015-072595). Technical support by Inmaculada Martinez (ICMM) is acknowledged. Additionally, Vignaswaran K. Veerapandiyan and Marco Deluca gratefully acknowledge support from the Austrian Science Fund (FWF): P29563-N36.